\documentclass[aps,prl,nofootinbib,twocolumn,superscriptaddress,preprintnumbers,floatfix]{revtex4-2} 

\usepackage[nodisplayskipstretch]{setspace}
\usepackage{cancel}
\usepackage{mathtools}

\usepackage{amssymb}
\usepackage{amsmath}
\usepackage{epsfig}
\usepackage{hyperref}
\usepackage{breakurl}
\usepackage{bm}
\usepackage{color}
\usepackage{slashed}
\usepackage{tikz}
\usetikzlibrary{decorations.markings}
\usepackage[english]{babel}
\DeclareMathOperator{\arccosh}{arccosh}

\newcommand\beq{\begin{equation}}
\newcommand\eeq{\end{equation}}
\def\bea{\begin{eqnarray}}
\def\eea{\end{eqnarray}}

\DeclareRobustCommand{\SkipTocEntry}[4]{}

\newcommand{\nn}{\nonumber}

\newcommand\beal{\begin{aligned}}
\newcommand\eeal{\end{aligned}}

\usepackage{xcolor}

\newcommand{\bel}{{\boldsymbol \ell}}

\newcommand{\bk}{{\boldsymbol k}}

\newcommand{\bp}{{\boldsymbol p}}

\newcommand{\bq}{{\boldsymbol q}}


\begin{document}

\preprint{DESY\, 24-029\\\phantom{~}}
\title{Local-in-time Conservative Binary Dynamics at Fourth Post-Minkowskian Order} 
\author{Christoph Dlapa}
\affiliation{ Deutsches Elektronen-Synchrotron DESY, Notkestr. 85, 22607 Hamburg, Germany}

\author{Gregor K\"alin}
\affiliation{ Deutsches Elektronen-Synchrotron DESY, Notkestr. 85, 22607 Hamburg, Germany}

\author{Zhengwen Liu}
\affiliation{Niels Bohr International Academy, Niels Bohr Institute, Blegdamsvej 17, 2100 Copenhagen, Denmark}
\affiliation{School of Physics and Shing-Tung Yau Center, Southeast University, Nanjing 210018, China}

\author{Rafael A. Porto}
\affiliation{ Deutsches Elektronen-Synchrotron DESY, Notkestr. 85, 22607 Hamburg, Germany}

\begin{abstract}
Leveraging scattering information to describe binary systems in generic orbits requires identifying local- and nonlocal-in-time tail effects. We report here the derivation of the universal (non-spinning) local-in-time conservative dynamics at fourth Post-Minkowskian order, i.e. ${\cal O}(G^4)$. This is  achieved~by computing the nonlocal-in-time contribution to the deflection angle, and removing it from the full~conservative value in \cite{4pmeft2,4pmeftot}. Unlike the total result, the integration problem involves two scales---velocity and mass ratio---and features multiple polylogarithms, complete elliptic and iterated elliptic integrals, notably in the mass ratio. We reconstruct the local radial action, center-of-mass momentum and Hamiltonian, as well as the exact logarithmic-dependent part(s),~all~valid~for generic orbits.  We incorporate the remaining nonlocal terms for elliptic-like motion to sixth Post-Newtonian order. The combined Hamiltonian is in perfect agreement in the overlap with the Post-Newtonian state of the art. The results presented here provide the most accurate description of gravitationally-bound binaries harnessing scattering data to date, readily applicable to waveform~modelling.  
\end{abstract}
\maketitle

{\bf Introduction.}~Motivated by the impending era of high-precision gravitational-wave (GW) astronomy with observatories such as LISA \cite{LISA}, the Einstein Telescope \cite{ET} and the Cosmic Explorer \cite{CE}, and the incredibly rich amount of information expected from compact binary sources \cite{buosathya,tune,music,Maggiore:2019uih,Barausse:2020rsu,Bernitt:2022aoa}, the (long dormant \cite{west2}) Post-Minkowskian (PM) expansion in general relativity---entailing a perturbative series in $G$ (Newton's constant) but to all orders in the relative velocity---has experienced a resurgence in recent years, e.g. \cite{ira1,damour1,damour2,damour3,Goldberger:2016iau,cheung, donal, zvi1,paper1,paper2,b2b3,parra,pmeft,3pmeft,tidaleft,janmogul,pmefts,Jakobsen:2021zvh,Mougiakakos:2021ckm,Gabriele2,eftrad,4pmeft,4pmeft2,4pmeftot,4pmzvi,4pmzvi2,Jakobsen:2022psy,Jakobsen:2022zsx,Jakobsen:2023hig,Jakobsen:2023ndj,Damgaard:2023ttc,dklp,Frellesvig:2023bbf,cy,Buonanno:2024vkx,Driesse:2024xad}. This is, in part, thanks to the repurposing of modern integration techniques from collider physics (see \cite{dklp,cy} and references therein), which have led to a plethora of new results.~Notably, using worldline effective field theory (EFT) methodologies \cite{nrgr,nrgrs,iragrg,review,Goldberger:2022ebt}, the rapidly evolving state of the art includes the total relativistic impulse (yielding the scattering angle and emitted GW flux) of non-spinning \cite{4pmeft2,4pmeftot} and spinning \cite{Jakobsen:2023hig,Jakobsen:2023ndj} bodies to ${\cal O}(G^4)$, akin of a `three-loop' calculation in particle~physics, as well as partial results in the conservative sector at 5PM~\cite{Driesse:2024xad}.\vskip 4pt   

The derivations in \cite{4pmeft2,4pmeftot}, together with a (Firsov-type \cite{firsov,Landau:1969}) resummation scheme~\cite{paper1,paper2}, have led to an unprecedented agreement between analytic results and numerical simulations \cite{Damour:2022ybd,Hopper:2022rwo,Rettegno:2023ghr}, paving the way to more accurate waveform models for hyperbolic encounters. However, due to nonlocal-in-time effects~\cite{Damour:2014jta,tail}, unbound results cannot be used to describe generic elliptic-like motion (away from the large-eccentricity limit~\cite{4pmeft2}). As~shown in~\cite{b2b3}, the binding energy for quasi-circular orbits obtained from scattering results (via the ``boundary-to-bound" (B2B) analytically continuation \cite{paper1,paper2}) does not reproduce---other than logarithms---the known Post-Newtonian (PN) values \cite{Damour:2014jta,tail,Marchand:2017pir,Foffa:2019rdf,nrgr4pn2} (see also \cite{Khalil:2022ylj}). Hence, to fully harness the power of scattering  calculations, a separation between local- and nonlocal-in-time effects was thus imperative. In this letter we report the derivation of the universal (non-spinning) local-in-time conservative dynamics of binary systems at ${\cal O}(G^4)$. This is obtained via a direct computation of~the nonlocal-in-time  contribution to the scattering angle. Following \cite{b2b3,binidam2}, the calculation entails an integral over the energy spectrum times the logarithm of the center-of-mass GW frequency. To solve the integration problem, we implement the methodology of differential equations, already used in \cite{4pmeft2,4pmeftot}.  However, unlike the total impulse, which obeys a simple (power-law) mass scaling \cite{paper1,damour3}, isolating (non)local effects in a gauge-invariant fashion entails dealing with two relevant scales: the velocity and mass ratio. Despite the complexity of the two-scale problem, we find that it can be factorized into solving two second-order Picard-Fuchs (PF) equations. The nonlocal part of the angle features multiple polylogarithms (MPLs), complete elliptic integrals, and integrations thereof. We find agreement in the overlap with the 6PN values in~\cite{binidam2}.\vskip 4pt

We derive the local-in-time contribution to the conservative scattering angle by removing the (unbound)~nonlocal terms from the  total result in \cite{4pmeft2,4pmeftot}. The~local radial action follows directly via the B2B map \cite{paper1,paper2,b2b3}.~Using the relations in \cite{paper1,paper2}, we reconstruct the universal local-in-time center-of-mass momentum and Hamiltonian in isotropic gauge, together with the complete logarithmic dependence, all applicable to generic motion. We also provide---for all practical purposes---results expanded to 30 orders in the (symmetric) mass ratio and all orders in the velocity (with an error beyond 30PN). To incorporate the remaining (non-logarithmic) nonlocal part of the bound dynamics, we adapt to our isotropic gauge the values obtained in \cite{binidam2} to 6PN order. The combined Hamiltonian at ${\cal O}(G^4)$ perfectly matches in the overlap with the state of the art in PN theory~\cite{binidam1,binidam2,Khalil:2022ylj}. The results presented here can be directly inputed onto waveforms models for gravitationally-bound eccentric orbits, potentially increasing their accuracy by incorporating an infinite tower of (local-in-time) velocity corrections.\vskip 4pt

 {\bf (Non)local-in-time tail effects.}  The scattering of the emitted radiation off of the binary's gravitational potential, or ``tail effect", enters in the 4PM conservative dynamics both through local- as well as nonlocal-in-time interactions \cite{Damour:2014jta,tail, Marchand:2017pir,Foffa:2019rdf,nrgr4pn2}. Because of this, although an effectively local description is possible to any order \cite{4pmeft2,4pmeftot}, the coefficients of the radial action (or Hamiltonian) depend on the type of motion, and therefore are not related via analytic continuation for generic orbits. Our strategy is to identify the local- and nonlocal-in-time parts of ${\cal S}_r$, the total radial action. Due to the structure of tail effects \cite{Damour:2014jta,tail,b2b3}, the nonlocal-in-time tail terms can be shown to take the gauge-invariant form 
\beq
 {\cal S}^{(\rm nloc)}_{r\,} = -\frac{G E}{2\pi} \int_{\omega} \frac{dE}{d\omega}  \log \left(\frac{4\omega^2}{\mu^2} e^{2\gamma_E}\right)\,,\label{nloc1}
\eeq   
where $\int_{\omega} \equiv \int_{-\infty}^{+\infty} \frac{d\omega}{2\pi}$, $E$ and $\frac{dE}{d\omega}$ are the total (incoming) energy and emitted GW spectrum in the center-of-mass frame. The `renormalization scale' $\mu$, which cancels against a similar term in the local-in-time part \cite{4pmeft,4pmeft2,tail}, can be arbitrarily chosen. The  factor of $4e^{2\gamma_E}$ ( with $\gamma_E$ Euler's constant)  follows the PN conventions \cite{Damour:2014jta,tail}. An explicit derivation of  \eqref{nloc1} in the context of the PM expansion can be found in \cite{b2b3}, see also \cite{binidam2} for a discussion in the PN regime. \vskip 4pt For unbound motion, the scattering angle is given by $\frac{\chi}{2\pi} = -\partial_j   {\cal I}_r$ , with ${\cal I}_r \equiv \frac{{\cal S}_r}{GM^2\nu}$ and $j \equiv \frac{J}{GM^2\nu}$ the (reduced) radial action and angular momentum, and $M=m_1+m_2$, $q=m_2/m_1$ ($m_2 \leq m_1$), $\nu =m_1m_2/M^2$  the total mass, mass ratio, and symmetric mass ratio, respectively.  We split the PM coefficients of the deflection angle in impact parameter space as,
\beq \frac{\chi}{2} = \sum_{n=1} \left(\chi_b^{(n)} + \chi_b^{(n)\log} \log \frac{\mu b}{\Gamma}\right) \left(\frac{GM}{b}\right)^n,\eeq 
where $\gamma \equiv u_1\cdot u_2$ (using the mostly negative metric convention), $u_{1,2}$ the incoming velocities, and $\Gamma \equiv E/M = \sqrt{1+2\nu(\gamma-1)}$. (The reader should keep in mind that logarithms of the velocity may still appear in both coefficients.) In the remaining of this letter we choose $\mu \equiv 1/GM$ for the renormalization scale.
 
\vskip 4pt {\bf Integrand construction.} Because of the overall factor of $G$ in \eqref{nloc1}, it is sufficient to construct the  integrand to ${\cal O}(G^3)$. Using the results in \cite{eftrad}, and multiplying by a factor of $(2e^{\gamma_E}k\cdot u_{\rm com})^{2\tilde\epsilon}$, with $k=(\omega,\bk)$ the (on-shell) radiated momentum, and $u_{\rm com} \equiv \frac{(m_1 u_1+m_2u_2)}{E}$ the center-of-mass velocity, we readily derive a covariant version which, after projecting on the center-of-mass frame, matches at  ${\cal O}(\tilde\epsilon)$ the expression in \eqref{nloc1}. We~find it  convenient to distinguish the $\tilde \epsilon$-expansion from the standard $D=4-2\epsilon$ that we use for dimensional regularization. To the families of  (two-loop) scalar integrals introduced in \cite{eftrad} for computing the total impulse, 
we add,
\beq
  \begin{aligned}
    I^{\pm\pm;T_5\dots T_9}_{\nu_1\dots\nu_{10}}&=\int_{\ell_1,\ell_2}\frac{{\delta}^{(\nu_1-1)}(\ell_1\cdot u_1){\delta}^{(\nu_2-1)}(\ell_2\cdot u_2)}{(\pm \ell_1\cdot u_2)^{\nu_3} (\pm \ell_2\cdot u_1)^{\nu_4}} \\
    &\qquad\qquad\times(k\cdot u_{\textrm{com}})^{2\tilde\epsilon-\nu_{10}}\prod_{j=5}^9 \frac{1}{D_{j,T_j}^{\nu_j}}\,,
  \end{aligned}
\eeq
with a noninteger powered propagator, where we use the same notation as in \cite{eftrad} (see also \cite{dklp}). The radiative momentum is rewritten as $k^\alpha=\ell^\alpha_1+\ell^\alpha_2-q^\alpha$, with $q^\alpha \equiv (q^0,\bq)$ the  momentum transfer, obeying $q\cdot u_a=0$ (not to be confused with the mass ratio). The choice of $i0^+$-prescription for the square propagators, either retarded or advanced, is encoded in $T_j\in\{\textrm{ret},\textrm{adv}\}$:
\bea 
    D_{5,\textrm{ret/adv}} &=& (\ell_1^0 \pm i0)^2 - \bel_1^2 \,, \nn
    D_{6,\textrm{ret/adv}} = (\ell_2^0 \pm i0)^2 - \bel_2^2 \,,\\
    D_{7,\textrm{ret/adv}} &=& (\ell_1^0 + \ell_2^0 + q^0 \pm i0)^2 - (\bel_1+\bel_2-\bq)^2 \nn \,,\\
    D_{8,\textrm{ret/adv}} &=& (\ell_1^0 -q^0 \pm i0)^2 - (\bel_1 - \bq)^2 \,,\nn \\
    D_{9,\textrm{ret/adv}} &=& (\ell_2^0 - q^0 \pm i0)^2 - (\bel_2 - \bq)^2\,.
\eea
Using integration-by-parts (IBP) reduction techniques implemented in the packages \texttt{LiteRed} \cite{Lee:2013mka} and \texttt{FiniteFlow} \cite{Peraro:2019svx}, we find 17 master integrals contributing to the radiation region (where the $k$ momentum goes on-shell), which isolates the contribution to the energy loss from the total impulse. It is possible to select integrals such that we can take $\nu_1=\nu_2=1$, $\nu_{10}=0$. The final set, specified by $\nu_{3\cdots 9}$, becomes 
\beq
\nn
\small
\begin{aligned}
  &(-1,0,0,0,1,1,1)\,, &
  &(-1,0,0,0,1,1,2)\,,&
  &(-1,0,0,0,1,2,1)\,, & \\ 
  &(-1,0,0,0,2,1,1)\,,&
  &(0,-1,0,0,1,1,1)\,, &
  &(0,-1,0,0,1,1,2)\,,& \\
  &(-1,0,0,1,1,0,1)\,, &
  &(0,-1,0,1,1,0,1)\,, &
  &(-1,0,0,1,1,1,1)\,,& \\
  &(-1,0,1,0,1,1,0)\,, &
  &(0,-1,1,0,1,1,0)\,,& 
  &(-1,0,1,0,1,1,1)\,,& \\
  &(-1,0,1,1,1,1,1)\,, &
  &(-1,0,1,1,1,1,2)\,,& 
  &(-1,0,1,1,2,1,1)\,,&\\
  &(0,1,1,1,1,0,0)\,, &
  &(1,0,0,0,1,1,1)\,,&
\end{aligned}
\eeq
modulo different choices of  $i0^+$-prescriptions ($T_{5\cdots 9}$) and signs in front of linear propagators.

\vskip 4pt {\bf Integration.} To solve for the master integrals, we derive differential equations in $x$ and the mass ratio, $q$, where $x$ is given by $\gamma=\frac{1}{2}\left(x+\frac{1}{x}\right)$. We then adopt the strategy of an $\epsilon$- (and $\tilde{\epsilon}$-)regular basis \cite{Lee:2019wwn}, such that we can set $\epsilon=0$, and consider the expansion of the integrand, differential equations, and boundary constants, only to $\mathcal{O}(\tilde{\epsilon})$. The latter are determined via a small-$q$ expansion, together with the techniques described in \cite{dklp} (adapted to the new factors of $\tilde{\epsilon}$).  From this setting, it is then straightforward to find a solution of the differential equations through iterated integration.
\vskip 4pt

For the parts containing MPLs, and similarly to the $x$ variable, it is useful to rationalize the square root of the energy $\frac{E}{m_1}=\sqrt{1+2\gamma q+q^2}=\sqrt{(q+x)\left(q+\frac{1}{x}\right)}$, by introducing a new variable, $y$, defined through ${q^{-1}=-\gamma-\frac{v_{\infty}}{2}\left(y+\frac{1}{y}\right)}$, with $v_{\infty} \equiv \sqrt{\gamma^2-1}$. Hence, we find the traditional harmonic polylogarithms with letters $\{x,1+x,1-x\}$ \cite{3pmeft,eftrad}, as well as MPLs~which depend on the velocity and mass ratio via the new letters:  ${\{y,1+y,1-y,y-\frac{1+x}{1-x},y-\frac{1-x}{1+x},1 + 2\frac{1-x}{1+x}y + y^2\} }$. In addition to MPLs, the solution to the differential equations depend on another set of functions, through an \emph{a priori} irreducible fourth-order PF equation, already at ${\cal O}(\tilde{\epsilon}^0)$. However, a Baikov representation \cite{Baikov:1996iu,Frellesvig:2017aai} of the maximal cut suggests a simpler Calabi-Yau two-fold as the relevant geometry. 
Indeed, in terms of the  variables $(q x, \frac{q}{x})$, the differential equations can be solved, in the first and subsequently the second variable, via two equivalent second-order PF equations (per variable). The solution can then be written in terms of products of $\mathrm{K}$'s (such as the $f_1$ in \eqref{eq:elliptic-kernels} below) as well as the leading three derivatives w.r.t.~the mass ratio. As in previous PM computations, e.g. \cite{4pmeft,4pmeft2,4pmeftot}, $\mathrm{K}(z) = \int_0^1 \frac{dt}{\sqrt{(1-t^2)(1-zt^2)}}$, is the complete elliptic integral of the first kind.\vskip 4pt

After the leading order solution is known, it is then straightforward to obtain the $\mathcal{O}(\tilde{\epsilon})$ part. We find that it can be written in terms of (at most) two-fold iterated integrals, with elliptic kernels depending on the mass ratio, $q$, as the integration variable. The full set is given by:
 \begin{align}
   f_{1}&=\frac{\mathrm{K}(-q x) \mathrm{K}\left(1+ \frac{q}{x}\right) - 
  \mathrm{K}\left(- \frac{q}{x}\right) \mathrm{K}(1 + q x)}{\pi},\qquad\nn \\  
  f_{2}&=\frac{f_{1}}{q}\,, \quad\quad f_{3}=\partial_x f_{1}\,,\quad\quad  f_{4}=\frac{\partial_x f_{1}}{q}\,, \label{eq:elliptic-kernels}
\\
  f_{5}&=\left[\frac{1-x^2}{x}(1+ q\,\partial_q)-\frac{1-q^2}{q}x\,\partial_x\right]\frac{f_{1}}{\sqrt{(q+x)\left(q+\frac{1}{x}\right)}}\nn.
  \end{align}
Remarkably, while individually this is not the case, the combination of complete elliptic integrals in $f_1$ has a simple power-series expansion in the PN limit ($x\to1$). Furthermore, the $f_i$'s are real, and have (at most) simple poles in~$q$. Let us point out, however, that a simplified version of the iterated integrals may still be possible. In particular, upon assigning to $\mathrm{K}(z)$ a transcendental weight {\it one}, we notice that the iterated integrals would have up to weight {\it four},  in contrast to the MPL part with maximum weight {\it two}. Hence, we expect that either the na\"ive assignment is incorrect or an even simpler form exists. We leave this open for future work. 

\vskip 4pt {\bf Scattering angle.} After solving for the master integrals and plugging them back into the integrand, we arrive at the radial action, and from there to the   nonlocal-in-time contribution to the deflection angle, $\chi_{b (\rm nloc)}^{(4)}$ and $\chi_{b (\rm nloc)}^{(4)\log}$,  at 4PM order. As anticipated in \cite{4pmeft}, the logarithmic part takes on a simple closed form,
\begin{align}
\label{bnloclog}
&  \frac{1}{\pi\Gamma}\chi_{b (\rm nloc)}^{(4)\log} = -2 \nu \chi_{2\epsilon}(\gamma)=\\
  &\frac{-2\nu}{{\left(\gamma ^2-1\right)^2}}
    \left(h_5
    +h_{9} \log \left(\frac{\gamma +1}{2}\right)
    +\frac{h_{10} \arccosh(\gamma )}{\sqrt{\gamma ^2-1}}
  \right)\,,\nn
\end{align}
with $\chi_{2\epsilon}$ introduced in \cite{4pmeft}, and the $h_{5,9,10}$ are polynomials depending only on $\gamma$, which also enter the non-logarithmic part (see below). The latter, on the other hand, also involves a set of iterated integrals in the mass ratio. Despite its complexity, it is straightforward to construct a ``self-force" (SF) expanded version, for which we find the generic form,  valid to any $n$SF order, 
\begin{widetext}
  \beq\label{chinlocnlog}
  \begin{aligned}
    \frac{1}{\pi\Gamma}\chi_{b (\rm nloc)}^{(4) (n\rm SF)} &=  \frac{\nu}{\left(\gamma ^2-1\right)^2}\left\{
    h_1
    +\frac{\pi ^2 h_2}{\sqrt{\gamma ^2-1}}
    +h_3 \log \left(\frac{\gamma +1}{2}\right)
    +\frac{h_4 \arccosh(\gamma )}{\sqrt{\gamma ^2-1}}
    +h_5 \log \left(\frac{\gamma -1}{8}\right)
    +h_6 \log ^2\left(\frac{\gamma   +1}{2}\right)\right.\\
    &\quad
    +h_7 \arccosh(\gamma )^2
    +\frac{h_8 \log (2) \arccosh(\gamma )}{\sqrt{\gamma ^2-1}}
    +h_9 \log \left(\frac{\gamma -1}{8}\right) \log \left(\frac{\gamma +1}{2}\right)\\
    &\quad\left.
    +\frac{h_{10} \log \left(\frac{\gamma ^2-1}{16}\right) \arccosh(\gamma )}{\sqrt{\gamma ^2-1}}
    +h_{11} \text{Li}_2\left(\frac{\gamma -1}{\gamma +1}\right)
    +\frac{h_{12} \left[\arccosh(\gamma )^2+4 \text{Li}_2\left(\sqrt{\gamma^2-1}-\gamma \right)\right]}{\sqrt{\gamma ^2-1}}
    \right\}\,.
  \end{aligned}
  \eeq
\end{widetext}
Due to the structure of the full solution, except for the $h_1$, $h_3$ and $h_4$ carrying information from the (iterated) elliptic sector ($h_{1,4}$) and the new letters in the MPLs depending on the mass ratio ($h_3$), the remaining $h_i$'s are SF-exact. We find the $n$SF coefficients may be split as \beq h_i = h_i^{(0)}(\gamma) + \sqrt{1-4\nu}\, h_i^{(1)}(\gamma) + \Delta h_i (\gamma,\nu)\,,\eeq where the $h_i^{(0)}(\gamma)$, $h_i^{(1)}(\gamma)$, are polynomials in $\gamma$ only. The $\Delta h_i (\gamma,\nu)$ vanish except when $i=1,3,4$, for which they become polynomials both in $\gamma$ and~$\nu$, up to ${\cal O}(\nu^n)$. We~provide in an ancillary file their values up to $n=30$. The 30SF result (with an error beyond 30PN) is in perfect agreement in the overlap with the 6PN values in~\cite{binidam2}.\vskip 4pt Let us emphasize that the definition of nonlocal-in-time in \cite{binidam1,binidam2} includes not only the expression in \eqref{nloc1} ($W_1$ in \cite{binidam2}), but also an extra contribution ($W_2$ in \cite{binidam2}). Due to the local-in-time (and gauge-dependent) nature of $W_2$, we do not add it to~\eqref{nloc1}. Therefore, \eqref{chinlocnlog} agrees (in the overlapping realm of validity) with the scattering angle obtained from the $W_1$-only terms in Eq.~(3.14) of~\cite{binidam2}.\vskip 4pt After subtracting from the total conservative angle in \cite{4pmeft2,4pmeftot}, we arrive at the local-in-time counterpart,\footnote{Although amenable to a conservative-like description of the relative dynamics, we keep the other (time-symmetric) radiation-reaction corrections, i.e.~``2rad" in \cite{4pmeftot}, in the dissipative part.} 
\beq
\chi_{b (\rm loc)}^{(4)} = \chi_{b \rm (tot)}^{(4)\rm cons} - \chi_{b (\rm nloc)}^{(4)}\,,\quad
\chi_{b (\rm loc)}^{(4)\log} = - \chi_{b (\rm nloc)}^{(4) \log}\,,\label{chiloc}
\eeq 
where we used the fact that the $\log \frac{\mu b}{\Gamma}$ cancels out in the total value. The result in \eqref{chiloc} can now be used to describe generic bound orbits, as we discuss next.\vskip 4pt 

{\bf Local-in-time conservative dynamics.} Following the B2B dictionary, the local-in-time (reduced) bound radial action takes the form \cite{paper1}
\beq
i^{4 \rm PM}_{r (\rm loc)} = \frac{2v_{\infty}^4}{3(\Gamma j)^3} \left(\frac{\chi_{b (\rm loc)}^{(4)}}{\pi \Gamma}+\frac{\chi_{b (\rm loc)}^{(4)\log }}{2\pi \Gamma} \log \frac{j^2}{v_{\infty}^2}\right)\,.
\eeq
 Using the expressions in \cite{paper1,paper2}, and a dimensionally rescaled distance $\hat r = r/(GM)$, we can also reconstruct the center-of-mass momentum (notice we use different conventions w.r.t. \cite{paper1,paper2}),
\beq
{\hat \bp}^2 = \frac{v_{\infty}^2}{\Gamma^2} \left( 1 + \sum_{n=1} \frac{1}{\hat r^n} \left( f_{n}  + f_{n}^{\log} \log \hat r \right)\right) \,,
\eeq
with $\hat \bp \equiv \bp/(M\nu)$, and Hamiltonian, $\hat H \equiv H/(M\nu)$, 
\beq
\label{H}
\hat H =  \hat E+ \sum_{i=1} \frac{1}{\hat r^i} \left( \hat c_{i}+ \hat c_{i}^{\log}  \log \hat r\right)\,,
\eeq
where $\hat E = \sum_a \hat E_a$, 
$\hat E_a = \sqrt{\hat \bp^2+\left(\frac{m_a}{M\nu}\right)^2}$.\vskip 4pt
The coefficients $\big(\hat c_{4 (\rm loc)},\hat c^{\log}_{4 \rm (loc)} \big)$ are displayed in ancillary files. \vskip 4pt

{\bf Universal logarithms.} Nonlocal-in-time tail effects also contribute with a $\log \hat r$ term in the bound dynamics. Performing a small-eccentricity expansion of~\eqref{nloc1}, and using Kepler's law ($\log \Omega = -\frac{3}{2} \log \hat r + \cdots $, with $\Omega$ the 1PM orbital frequency), we find  \beq \frac{\hat c_{4(\rm nloc)}^{\log}}{\hat r^4} = -\frac{3}{2} \frac{\hat c_{4(\rm loc)}^{ \log}}{\hat r^4}= -3 G \frac{\Gamma}{\nu} \left.\frac{dE}{dt}\right|_{\rm 3PM} \,,\eeq  
where \big($\xi\equiv \frac{\hat E_1\hat E_2}{(\hat E_1+\hat E_2)^2}, \gamma = \nu (\hat E_1\hat E_2 + \hat \bp^2)$\big) 
\beq
\label{de3pm}
G \left.\frac{dE}{dt}\right|_{\rm 3PM}(\hat r,\hat \bp^2)= - \frac{4\nu^3}{3{\hat r}^4} \frac{\gamma^2-1}{\Gamma^3\xi}\chi_{2\epsilon}(\gamma)\,,
\eeq
is the energy flux at 3PM order \cite{4pmeft,b2b3}. Similarly, 
\bea
 f_{4(\rm nloc)}^{\log} &=&  -\frac{3}{2} f_{4(\rm loc)}^{ \log}  
 = -8\Gamma \nu \,\chi_{2\epsilon}\,,
\eea
consistently with \eqref{bnloclog}. Hence, adding both terms,
\bea
\hat H^{{\rm ell (\log) }}_{\rm 4PM} \label{Hlog} 
&=&  \frac{4 \nu^2}{3\hat r^4}\frac{(\gamma^2-1)}{\Gamma^2\xi}\chi_{2\epsilon} \log \hat r\,, \eea
 and likewise,
 \bea
\left(\hat \bp^2\right)^{{\rm ell (\log) }}_{\rm 4PM} \label{plog} 
&=&  - \frac{8\nu v_{\infty}^2 }{3\Gamma \hat r^4}   \,\chi_{2\epsilon} \log \hat r\,, \eea
for the full logarithmic dependence of the bound Hamiltonian and center-of-mass momentum at 4PM.\vskip 4pt 

{\bf Towards the complete bound dynamics.}  Putting together the local-in-time coefficient plus exact logarithmic part, the total bound Hamiltonian up to 4PM order may be written as
\bea
\label{Htot4pm} \hat H^{\rm ell}_{\rm 4PM}  &=&  \sum_{i=1}^{i=4} \frac{\hat c_{i\rm (loc)}}{\hat r^i}  + \sum_{i=1}^{i=4}  \frac{\hat c_{i\rm (nloc)}}{\hat r^i}  \\ &+&  \frac{4 \nu^2}{3\hat r^4}\frac{(\gamma^2-1)}{\Gamma^2\xi}\chi_{2\epsilon} \log \left(\frac{\hat r}{e^{2\gamma_E}}\right) \,,  \nn 
 \eea
where we have absorbed the factor of $e^{2\gamma_E}$ that arises from \eqref{nloc1} into the logarithm. The $\hat c_{1|2|3(\rm loc)}$ are the known local-in-time PM coefficients up to 3PM order \cite{cheung,pmeft,zvi1,3pmeft}, and $\hat c_{4(\rm loc)}$ is reported here for the first time. \vskip 4pt To complete the knowledge of the bound dynamics, we are still missing the \big[non-$\log \left(\frac{\hat r}{ e^{2\gamma_E}}\right)$\big] nonlocal-in-time contributions, $\hat c_{i\rm (nloc)}$, which depend on the trajectory. These are more difficult to compute in a PM scheme, since they are often needed in the opposite limit of quasi-circular orbits, thus entering at all PM orders!  Yet,~they can be readily obtained within the PN approximation by evaluating the radial action in~\eqref{nloc1} in a small-eccentricity expansion. Adapting the ($W_1$-only) results in~\cite{binidam2} to the isotropic gauge, we quote their values in the supplemental material to 6PN and eight order in the eccentricity.~The combined Hamiltonian in \eqref{Htot4pm} is in perfect agreement to ${\cal O}(G^4\hat \bp^6)$ with the $\hat H^{\rm ell}_{\rm 6PN(4PM)}$ derived~in~\cite{Khalil:2022ylj} using the state of the art in PN theory, while at the same time it incorporates all-order-in-velocity corrections. Ready-to-use expressions for the full results and 30SF-approximate are collected in ancillary files.\vskip 4pt

\vskip 4pt {\bf Conclusions.} Novel integration techniques in combination with EFT methodologies have been extremely successful in reaching the very state of the art in our understanding of scattering dynamics in general relativity, including conservative and dissipative effects \cite{dklp,Driesse:2024xad}. However, as illustrated in \cite{b2b3,Khalil:2022ylj}, although local-in-time and logarithms are universal, the full hyperbolic results fail to describe quasi-circular binaries. This is due to the presence of orbit-dependent (non-logarithmic) nonlocal-in-time effects, which preclude a smooth analytic continuation via the B2B map \cite{paper1,paper2}. Hence, up until now, we were lacking a direct correspondence to generic bound motion, notably for the conservative sector.\vskip 4pt We have computed the nonlocal-in-time contribution to the deflection angle, and removed it from the total conservative value in \cite{4pmeft2,4pmeftot}, thus yielding the local-in-time counterpart. We then derived the radial action, center-of-mass (isotropic-gauge) momentum and Hamiltonian, as well as the total logarithmic-dependent part(s), all applicable to generic motion. Upon adapting the (non-logarithmic) nonlocal-in-time effects for elliptic-like orbits computed in the PN expansion \cite{binidam2}, the combined total Hamiltonian becomes the most accurate description of gravitationally-bound binary systems obtained from PN/PM data to date, readily applicable to waveform modelling. Studies assessing the implications of our results towards constructing high-precision GW templates, as well as the derivation of a PM version of nonlocal-in-time effects for bound orbits, are underway.\vskip 4pt 

{\bf Acknowledgements}. We thank Massimiliano M. Riva for useful discussions. We thank Sebastian P\"ogel and  Yu Jiao Zhu for suggesting the Calabi-Yau two-fold as the relevant geometry and Matthias Wilhelm for useful discussions on the integration problem. We are grateful to Donato~Bini and Mohammed Khalil for their help adapting the nonlocal-in-time (elliptic-like) terms from $W_1$ in \cite{binidam2} into the isotropic~gauge. The work of CD, GK, and RAP  is supported by the ERC-CoG Precision Gravity: From the LHC to LISA provided by the European Research Council (ERC) under the European Union's H2020 research and innovation programme (grant No. 817791). ZL is supported partially by DFF grant 1026-00077B, the Carlsberg foundation, and the European Union's Horizon 2020 research and innovation program under the Marie Sklodowska-Curie grant agreement No. 847523 `INTERACTIONS'.

\bibliographystyle{apsrev4-1}
 \bibliography{ref4PM}
 \appendix
 \newpage
 \begin{widetext}

 \section{Supplemental Material}
 
 An accurate hybrid description of elliptic-like orbits can be constructed by incorporating into the Hamiltonian in~\eqref{Htot4pm} the \big[non-$\log\left(\frac{\hat r}{e^{2\gamma_E}}\right)$\big] nonlocal-in-time ($W_1$-only) contributions computed in PN theory to 6PN and ${\cal O}(e^8)$ in the small-eccentricity expansion in \cite{binidam2}, yielding 
 \bea
\label{Htot}\hat H^{\rm ell}(\hat r, \bp^2,\nu) &=& \hat E + \sum_{i=1}^{i=4} \frac{\hat c_{i\rm (loc)}(\hat \bp^2,\nu)}{\hat r^i}  + \frac{4 \nu^2}{3\hat r^4}\frac{(\gamma^2-1)}{\Gamma^2\xi}\chi_{2\epsilon}(\gamma) \log \left(\frac{\hat r}{e^{2\gamma_E}}\right) \\ &+&
 \sum_{i=1}^{i=4}  \frac{1}{\hat r^i}  \left\{\hat c^{{\rm 6PN}(e^8)}_{i\rm (nloc)}(\hat \bp^2,\nu)+ {\cal O}\big(\hat\bp^{2(8-i)}\big)\right\}+
\frac{1}{\hat r^5}\left(\hat c^{{\rm 4PN}(e^8)}_{5(\rm loc+nloc)}(\nu) -\frac{22\nu}{15} \log \left(\frac{\hat r}{e^{2\gamma_E}}\right)\right) +{\cal O}\left(\frac{\hat \bp^2}{\hat r^5}\right)\nn\,, 
\eea
where we have included also the known (static) 4PN correction at ${\cal O}(G^5)$,
\bea
\hat c^{{\rm 4PN}(e^8)}_{5(\rm loc+nloc)} &=& -\frac{1}{16} + \nu^2 \left(\frac{287\pi^2}{128}-\frac{1849}{48}\right)-\frac{5\nu^3}{8} \\
&+& \nu\left(\frac{76645\pi^2}{12288}-\frac{334259}{720}+\frac{5421492}{5}\log(2)+\frac{631071}{20}\log(3) - \frac{1953125}{4}\log(5) \right)\,,\nn
\eea
and
 \bea 
 \chi_{2\epsilon}(\gamma) &=&  -\frac{210 \gamma ^6-552 \gamma ^5+339 \gamma ^4-912 \gamma ^3+3148 \gamma ^2-3336 \gamma +1151}{32 (\gamma^2-1)^2}\\&+& \frac{3 \left(35 \gamma ^4+60 \gamma ^3-150 \gamma ^2+76 \gamma -5\right) }{16(\gamma^2 -1)}\log \left(\frac{\gamma+1}{2}\right)
    -\frac{3 \gamma  \left(2 \gamma ^2-3\right) \left(35 \gamma ^4-30 \gamma ^2+11\right) }{32 \left(\gamma ^2-1\right)^{2}}\frac{\arccosh(\gamma )}{\sqrt{\gamma^2-1}}\nn \\
 \gamma &=& \nu (\hat E_1\hat E_2 + \hat \bp^2)= \nu \hat\bp^2+\sqrt{1+\bp^2+\nu \hat\bp^2(\nu \hat\bp^2-2)}\,,\quad
  \Gamma^2\xi = \nu^2 \hat E_1 \hat E_2 = \nu \sqrt{1+\hat \bp^2+\nu \hat\bp^2(\nu \hat\bp^2-2)}\,.\nn
 \eea
 \vskip 12pt
For the remaining terms, translating from the results in \cite{binidam2} to our isotropic gauge, one finds
\begingroup
\allowdisplaybreaks
\begin{align}
  \hat c^{{\rm 6PN}(e^8)}_{1\rm (nloc)}(\hat{\bp}^2,\nu)&=
    \hat{\bp}^8 \nu\left[-\frac{17886}{175}+\frac{10834496 \log (2)}{45}+\frac{6591861 \log (3)}{700}-\frac{27734375 \log (5)}{252}\right]\\
    &\quad+\hat{\bp}^{10} \begin{multlined}[t]
      \left[\nu ^2 \left(-\frac{17886}{35}+\frac{10834496 \log (2)}{9}+\frac{6591861 \log (3)}{140}-\frac{138671875 \log (5)}{252}\right)\right.\\
      \left.+\nu  \left(\frac{8943}{175}-\frac{5417248 \log (2)}{45}-\frac{6591861 \log(3)}{1400}+\frac{27734375 \log (5)}{504}\right)\right]
    \end{multlined}\nn\\
    &\quad+\hat{\bp}^{12} \begin{multlined}[t]
      \left[\nu ^3 \left(-\frac{53658}{35}+\frac{10834496 \log (2)}{3}+\frac{19775583 \log (3)}{140}-\frac{138671875 \log (5)}{84}\right)\right.\\
      +\nu ^2 \left(\frac{80487}{175}-\frac{5417248 \log (2)}{5}-\frac{59326749 \log (3)}{1400}+\frac{27734375 \log (5)}{56}\right)\\
      \left.+\nu  \left(-\frac{26829}{700}+\frac{1354312\log (2)}{15}+\frac{19775583 \log (3)}{5600}-\frac{27734375 \log (5)}{672}\right)\right]\,,
    \end{multlined}\nn\\
  \hat c^{{\rm 6PN}(e^8)}_{2\rm (nloc)}(\hat{\bp}^2,\nu) &=
    \hat{\bp}^6\nu \left[\frac{578843}{1050}-\frac{21212984 \log (2)}{15}-\frac{282031389 \log (3)}{5600}+\frac{1296484375 \log (5)}{2016}\right]\\
    &\quad+\hat{\bp}^8 \begin{multlined}[t]
      \left[\nu ^2 \left(\frac{48747851}{13440}-\frac{2125906693 \log (2)}{378}+\frac{242787134673 \log (3)}{286720}+\frac{3671798828125 \log (5)}{1548288}\right.\right.\\
        \left.-\frac{96889010407 \log (7)}{221184}\right)+\nu  \left(\frac{81402259}{53760}-\frac{249145033 \log (2)}{60}-\frac{103980982797 \log(3)}{229376}\right.\\
        \left.\left.+\frac{12166079921875 \log (5)}{6193152}+\frac{96889010407 \log (7)}{884736}\right)\right]
    \end{multlined}\nn\\
    &\quad+\hat{\bp}^{10} \begin{multlined}[t]
      \left[\nu ^3 \left(\frac{1764244493}{134400}-\frac{50772177511 \log (2)}{3780}+\frac{15140243287719 \log (3)}{2867200}+\frac{15746212109375 \log (5)}{3096576}\right.\right.\\
      \left.-\frac{1065779114477 \log (7)}{442368}\right)+\nu ^2 \left(\frac{922629733}{107520}-\frac{188966394467 \log (2)}{7560}-\frac{7125985899279\log (3)}{2293760}\right.\\
      \left.+\frac{147239183828125 \log (5)}{12386304}+\frac{484445052035 \log (7)}{589824}\right)+\nu  \left(-\frac{28327409}{35840}+\frac{154094423 \log (2)}{72}\right.\\
      \left.\left.+\frac{527065116993 \log (3)}{2293760}-\frac{12533579921875 \log (5)}{12386304}-\frac{96889010407 \log (7)}{1769472}\right)\right]\,,
    \end{multlined}\nn\\
  \hat c^{{\rm 6PN}(e^8)}_{3\rm (nloc)}(\hat{\bp}^2,\nu) &=
    \hat{\bp}^4 \nu\left[-\frac{175727}{150}+\frac{233388968 \log (2)}{75}+\frac{16351713 \log(3)}{160}-\frac{405859375 \log (5)}{288}\right]\\
    &\quad+\hat{\bp}^6 \begin{multlined}[t]
      \left[\nu ^2 \left(-\frac{45488833}{4480}+\frac{105895904239 \log (2)}{13230}-\frac{10808816520303 \log (3)}{2007040}-\frac{28119126171875 \log (5)}{10838016}\right.\right.\\
        \left.+\frac{2588320706587 \log (7)}{1105920}\right)+\nu \left(-\frac{4548912331}{627200}+\frac{2922687496621 \log (2)}{132300}+\frac{13469503195629 \log (3)}{5734400}\right.\\
        \left.\left.-\frac{150369012359375 \log (5)}{14450688}-\frac{2588320706587 \log (7)}{4423680}\right)\right]
    \end{multlined}\nn\\
    &\quad+\hat{\bp}^8 \begin{multlined}[t]
      \left[\nu ^3 \left(-\frac{1108386017}{24192}+\frac{38790406370519 \log (2)}{1786050}-\frac{244047465883413 \log (3)}{10035200}\right.\right.\\
      \left.+\frac{1009279694921875 \log (5)}{877879296}+\frac{453841966033589 \log (7)}{89579520}\right)+\nu ^2 \left(-\frac{361592268899}{5644800}\right.\\
      \left.+\frac{116606471572979 \log(2)}{1071630}-\frac{448065058976289 \log (3)}{40140800}-\frac{181279182489765625 \log (5)}{3511517184}\right.\\
      \left.+\frac{3680972377512689 \log (7)}{358318080}\right)+\nu  \left(-\frac{3438069458633}{304819200}+\frac{369057536315537 \log (2)}{9185400}\right.\\
      \left.\left.+\frac{607401830370627 \log (3)}{80281600}-\frac{44240036362654375 \log (5)}{2341011456}-\frac{1267373911442149 \log (7)}{429981696}\right)\right]\,,
    \end{multlined}\nn\\
  \hat c^{{\rm 6PN}(e^8)}_{4\rm (nloc)}(\hat{\bp}^2,\nu) &=
    \hat{\bp}^2 \nu\left[\frac{27636}{25}-\frac{680106004 \log (2)}{225}-\frac{37122381 \log (3)}{400}+\frac{196484375 \log (5)}{144}\right]\\
    &\quad+\hat{\bp}^4 \begin{multlined}[t]
      \left[\nu ^2 \left(\frac{23663203}{1680}-\frac{9735062548 \log (2)}{33075}+\frac{14337306321183 \log(3)}{1254400}-\frac{2701666015625 \log (5)}{1354752}\right.\right.\\
      \left.-\frac{650540498447 \log (7)}{138240}\right)+\nu  \left(\frac{116416141}{9408}-\frac{526259559517 \log (2)}{13230}\right.\\
      \left.\left.-\frac{641012819877 \log (3)}{143360}+\frac{33874913921875 \log (5)}{1806336}+\frac{650540498447 \log(7)}{552960}\right)\right]
    \end{multlined}\nn\\
    &\quad+\hat{\bp}^6 \begin{multlined}[t]
      \left[\nu ^3 \left(\frac{10075896877}{120960}-\frac{2348423027149 \log (2)}{51030}+\frac{8674336284777 \log (3)}{286720}\right.\right.\\
      \left.-\frac{2232609748046875 \log (5)}{125411328}+\frac{250707235071713 \log (7)}{17915904}\right)+\nu ^2 \left(\frac{60245673427}{376320}\right.\\
      \left.-\frac{126132398166437 \log (2)}{1071630}+\frac{763693932388383 \log (3)}{8028160}+\frac{204623745011171875 \log (5)}{3511517184}\right.\\
      \left.-\frac{4304025048065071 \log (7)}{71663616}\right)+\nu  \left(\frac{102339488389}{2177280}-\frac{616925145960877 \log(2)}{3214890}\right.\\
      \left.\left.-\frac{144912376553769 \log (3)}{4014080}+\frac{52541416380715625 \log (5)}{585252864}+\frac{1554400159532395 \log (7)}{107495424}\right)\right]\,.
    \end{multlined}\nn
\end{align}
\endgroup
Upon PN expanding the total Hamiltonian in \eqref{Htot}, we find agreement with the expression for $\hat H^{\rm ell}_{\rm 6PN(4PM)}$ in \cite{Khalil:2022ylj}, in the overlapping realm of validity. 

 \vskip 4pt Let us conclude with a few remarks on the (non)local-in-time dynamics in \eqref{Htot}. In addition to $\gamma=u_1\cdot u_2$, the nonlocality {\it in time} introduces a new (mass-dependent) scale in the scattering problem, i.e. $u_{1(2)}\cdot u_{\rm com}$, yielding an intricate (non-algebraic) mass dependence for the nonlocal part (for fixed $\gamma$). Given the mass structure of the total result---guaranteed by the Lorentz invariance of the (unbound) radial action---this mass dependence transitions unscathed into the local-in-time (universal) contribution in \eqref{Htot}. One may then wonder, similarly to the hyperbolic case, whether the inclusion of a PM version of the $\hat c_{i(\rm nloc)}$'s for elliptic-like motion would also lead to a simpler structure for the total bound dynamics. After all, that is expected to be the case---through the B2B analytic continuation---in the large-eccentricity limit \cite{b2b3}.\footnote{Notice that $\gamma = 1+\frac{\Gamma^2-1}{2\nu}$, which introduces rather nontrivial mass dependence for fixed (binding) energy. Remarkably, when fully valid, the B2B map guarantees that adiabatic invariants for the bound problem can be written in terms of $\gamma$ and the angular momentum (up to conversion factors) \cite{paper1,paper2}.}  However, we would expect this to be manifest only in the full solution,\footnote{Modulo some subtleties, it is in principle possible to find an analytic continuation between the full (local+nonlocal) hyperbolic and elliptic {\it integrands} for the radial actions \cite{b2b3}, yielding e.g. the scattering angle and periastron advance, which would imply that (although not directly B2B-related for generic motion) the latter may still inherit the structure from the former.} and not for a given (PN/PM) truncation in the (opposite) regime of small eccentricities. For instance, due to the differences in the region of integration and energy spectrum (continuum vs. discrete), while nonlocal-in-time tail terms for hyperbolic-like motion first arise at ${\cal O}(G^4)$,  the above $\hat c_{i(\rm nloc)}$ coefficients for elliptic-like orbits enter at all orders in $G$ (starting at 4PN order). We will return to these issues elsewhere. 
 \end{widetext}
 
\end{document}